\documentclass[aps,amsfonts,amsmath,prd,preprint,nofootinbib]{revtex4}

\newcommand{\beq}{\begin{equation}}
\newcommand{\eeq}{\end{equation}}

\usepackage{epsfig,bbm}

\begin{document}
\title{Bubbles of Nothing in Flux Compactifications}

\author{Jose J.  Blanco-Pillado and Benjamin Shlaer}
\affiliation{Institute of Cosmology, Department of Physics and Astronomy,\\ 
Tufts University, Medford, MA 02155, USA}

\def\changenote#1{\footnote{\bf #1}}

\begin{abstract}
We construct a simple $AdS_4\times S^1$ flux compactification stabilized 
by a complex scalar field winding the single
extra dimension and demonstrate an instability to 
nucleation of a bubble of nothing.  This occurs when the Kaluza -- Klein
dimension degenerates to a point, defining the bubble surface.  
Because the extra dimension is stabilized by a flux, the 
bubble surface must be charged, in this case under the axionic part 
of the complex scalar.  This smooth geometry can be seen 
as a de Sitter topological defect with asymptotic 
behavior identical to the pure compactification.  We discuss 
how a similar construction can be implemented in more general 
Freund -- Rubin compactifications.
\end{abstract}

\maketitle
\thispagestyle{empty}
\section{Introduction}
\setcounter{page}{1}

Some time ago, Witten showed that the Kaluza -- Klein vacuum
suffers from a non-pertubative instability due to the semi-classical nucleation
of so-called {\it bubbles of nothing}.  Once formed, 
these smooth ``boundaries" of space expand and soon engulf 
the whole of spacetime \cite{Witten}.  Such instabilities
may be cause for concern regarding the viability of certain higher 
dimensional spacetimes as acceptable vacua.  It is therefore 
necessary to consider the existence of analogous decay processes 
for perturbatively stable compactifications.

Flux compactifications \cite{Douglas} provide an 
elegant solution to the moduli problem in higher dimensional 
field theories \cite{Cremmer, Freund, ME-6d,GellMann} as well 
as string theory \cite{GKP,KKLT}.  By 
stabilizing the extra dimensions with sufficiently high moduli 
masses, we can construct a model realistic enough to accommodate 
the low energy physics as well as a cosmological framework 
compatible with observations.  These masses are induced by a flux 
potential, which depends on a discrete set of flux winding 
numbers.  The diverse set of fluxes and associated winding numbers
gives rise to the multitude of (meta)stable vacua known as the 
string landscape \cite{Landscape}.  
Although supersymmetric flux compactifications
are known to be stable \cite{stable-SUGRA}, the more phenomenologically interesting 
compactifications may enjoy/suffer from several instabilities, 
including decompactification \cite{Giddings,BP-SP-V-1,CJR} or more general
transdimensional tunneling \cite{CJR, BP-SP-V-2}, flux-transitions 
\cite{Kachru,Frey, Freivogel, BP-SP-V-1,IS-Yang} and, as we will 
demonstrate here, nucleation of bubbles of nothing.%
\footnote{Bubbles of
nothing in flux compactifications have been recently discussed in \cite{IS-Yang}
by matching different spacetime geometries across a codimension {\em one} 
brane.  This is different from our present construction describing 
a geometry which is smooth everywhere, obviating the need to
postulate the existence of domain walls.}
The bubble 
of nothing geometry asymptotic to a given flux compactification
is a gravitational instanton which represents both the
decay mediator and subsequent classical evolution of the metastable 
spacetime.  This is the case for Witten's bubble of nothing in the minimal $5d$ 
Kaluza -- Klein (KK) model \cite{Witten}.

The outline of the paper is as follows.  In section II we review the
original bubble of nothing.  In section III we discuss how
these bubble solutions can be obtained as the spacetimes of 
de Sitter topological defects in a simple $5d$ flux compactification, 
where the extra dimension is stabilized by the
presence of a winding complex scalar field.  In section IV we
obtain numerical solutions describing these bubbles of nothing.
We conclude in section V and speculate on
similar constructions in more general flux compactifications.

\section{The Original Bubble of Nothing}
The bubble of nothing geometry introduced by Witten \cite{Witten} is easily obtained
from the five-dimensional Schwarzschild black hole,
\beq
ds^2 = -\left(1- {{\ell^2}\over {\rho^2}}\right) dt^2 + \left(1-
{{\ell^2}\over {\rho^2}}\right)^{-1} d\rho^2 + \rho^2 \left(d\psi^2 
+ \sin^2\psi \,\, d\Omega_2^2\right)~.
\eeq
It will be convenient for us to express this metric in terms
of a new radial coordinate, $r = \sqrt{\rho^2-\ell^2}$, as
\beq
ds^2 = -{{r^2}\over{r^2+\ell^2}} dt^2 + dr^2 + \left(r^2
+\ell^2\right) \left(d\psi^2 + \sin^2\psi\,\, d\Omega_2^2\right) \,\,.
\eeq
We can now Wick rotate two of the coordinates via
\beq
t \rightarrow i \ell y ~~~~~~\psi \rightarrow it + {{\pi}\over 2}
\eeq
to yield
\beq
\label{wittenmetric}
ds^2 = {{r^2}\over{1+ r^2/\ell^2}} dy^2 + dr^2 + \left(r^2 +
\ell^2\right) \left(-dt^2 + \cosh^2t\,\, d\Omega_2^2\right) \,\,.
\eeq
This is the bubble of nothing metric written in a somewhat
unfamiliar gauge.  (See the discussion in \cite{Garriga}.)  In the limit 
$r \rightarrow 0$, it becomes
\beq
ds^2 \approx r^2 dy^2 + dr^2 +  
\ell^2 \left(-dt^2 + \cosh^2t\,\, d\Omega_2^2\right)~,
\eeq
which is devoid of a conical singularity if we impose periodicity with
$0 \le y < 2 \pi$.  In this limit the $r-$slice degenerates onto 
a $2+1$ dimensional de Sitter space of size $\ell$, representing the induced metric
on the bubble surface.

In the limit of $r \gg \ell$, the metric asymptotes to
\beq
ds^2 \approx \ell^2 dy^2 + dr^2 +  r^2 \left(-dt^2 + \cosh^2t\,\,d\Omega_2^2\right)~,
\eeq
which represents the cartesian product of $4d$ Rindler space and a circle 
of circumference $2\pi \ell$.  The bubble of nothing geometry can therefore be
regarded as a deformation of Rindler space (times a circle), whereby 
the horizon region near $r \to 0$ is replaced with a smooth tip 
consisting of $dS_3\times {\cal B}_2$, where ${\cal B}_2$ is a cigar-shaped disk.  
This is illustrated in Fig.~\ref{confdiag1} below.
 
\begin{figure}[h] 
\centering
\includegraphics[width=3.0cm]{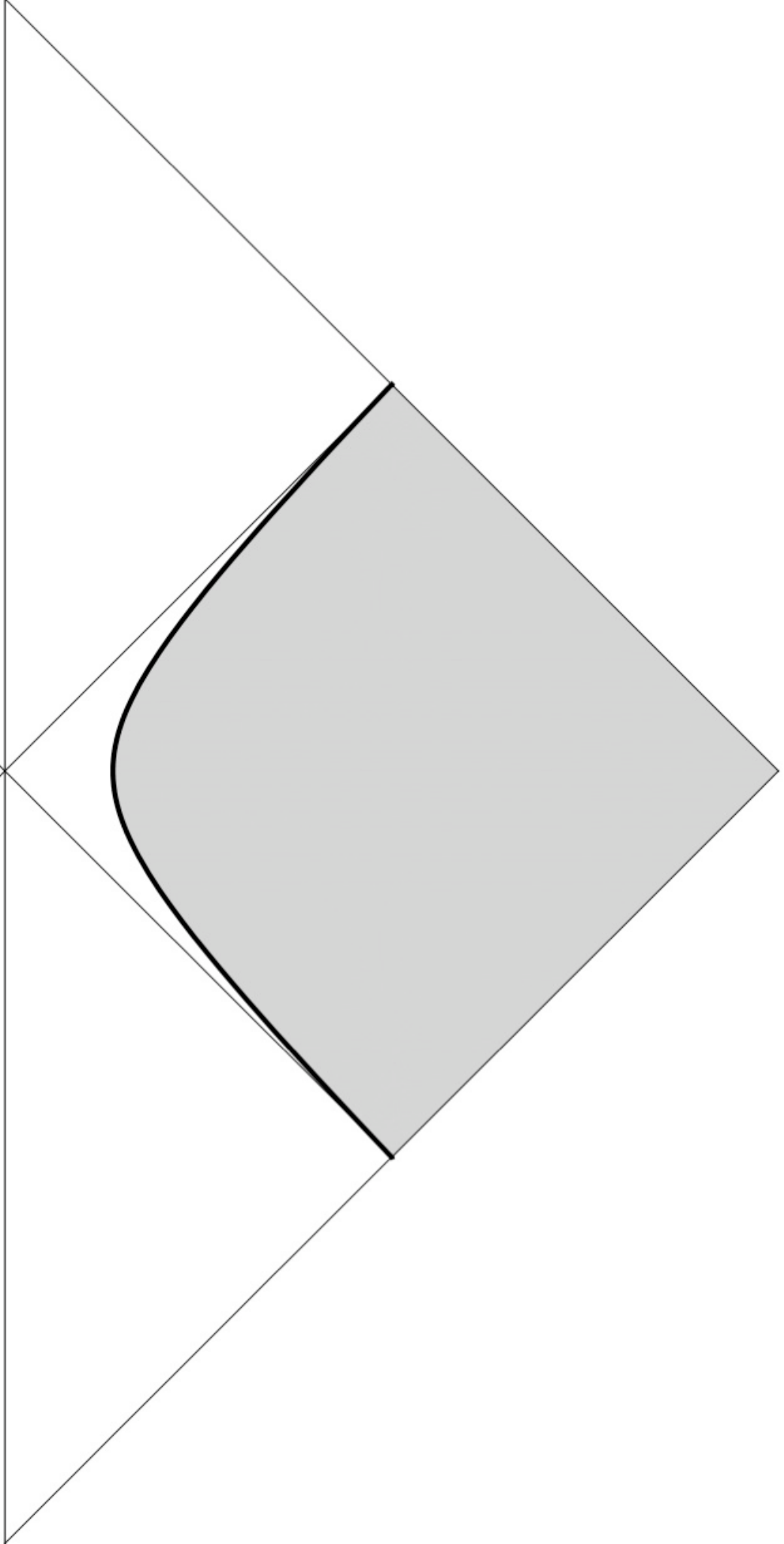}\hspace{2cm}
\includegraphics[width=8.0cm]{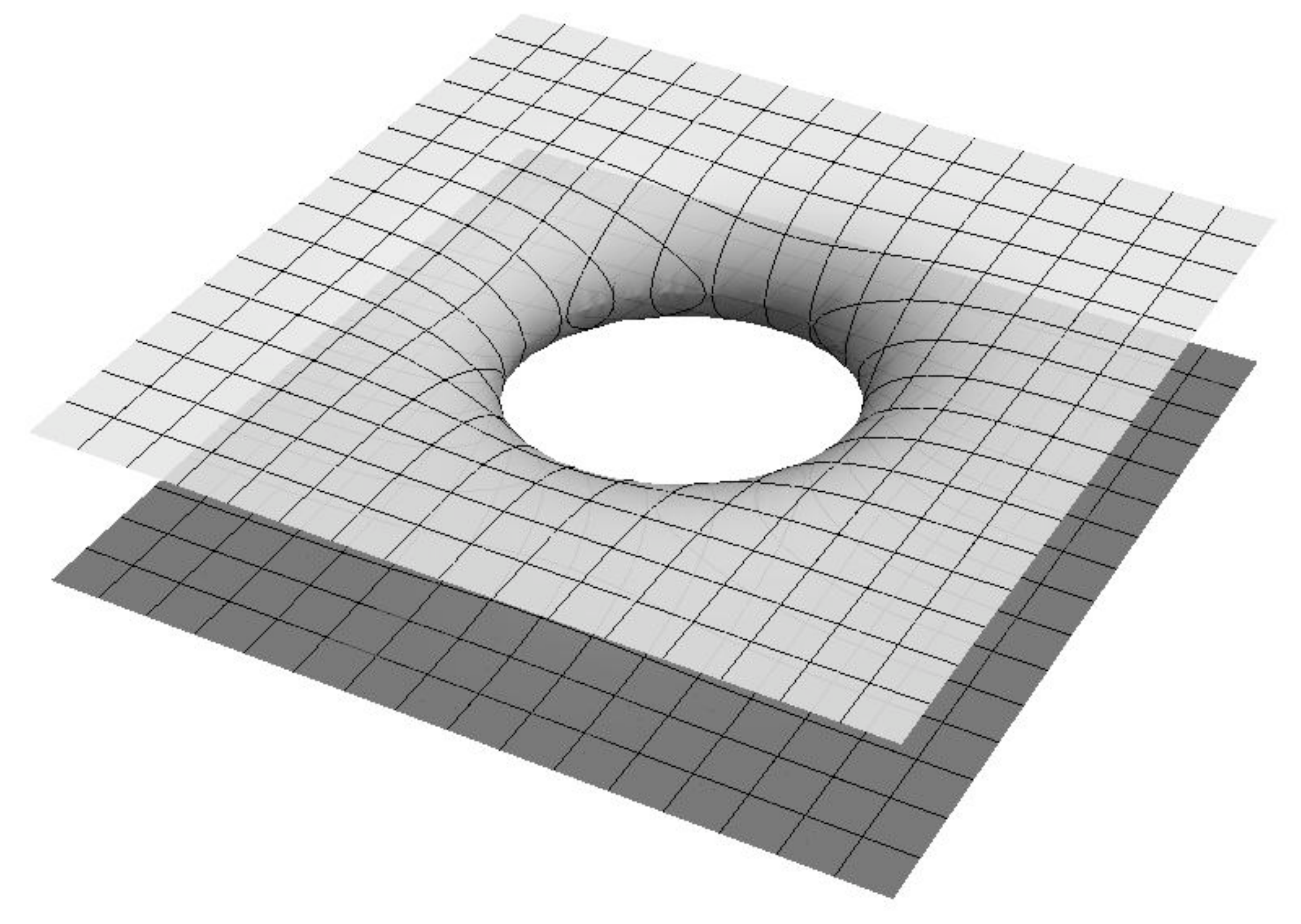}
\caption{Left:  A $4d$ conformal diagram for Witten's bubble of nothing.  
The spacetime only exists outside of the bubble (shaded region).  
Right: The bubble surface is smooth, and located where the
compactification volume, shown as the apparent vertical separation, degenerates to zero size.} 
\label{confdiag1}
\end{figure}

In order for this geometry to represent an instability of the
Kaluza -- Klein vacuum, it must have the same asymptotics as the pure 
compactification, in particular the same (zero) value of Schwarzschild mass.  
This can be seen to be the case in a number of ways.  The KK compactification possesses the 
eleven isometeries of Poincar\'e$\,\times\,U(1)$, which are broken down 
to the seven in $O(3,1)\times U(1)$ by the bubble of nothing (four
translations are lost).  This remaining symmetry is nevertheless 
larger than the five isometeries found in a generic
Schwarzschild -- Kaluza -- Klein spacetime: $O(3)\times {\mathbbm{R}}\times U(1)$, 
and so the mass of the bubble must be zero.  Hence, barring any symmetry 
in place to prevent bubble nucleation, the KK vacuum is unstable.

\section{Bubble of Nothing in a $5d$ Flux Compactification}

In this section we will discuss the construction
of a bubble of nothing geometry similar to that presented 
above, but with one crucial difference: we will stabilize the 
Kaluza -- Klein radion.  This is accomplished by introducing  
an axionic flux winding around the extra dimension.  The 
simplest example of this type of flux compactification 
was described in \cite{BP-SP-V-1} using the action for a complex scalar field
given by
\beq
S=\int{d^5 x \sqrt{- g} \left(  {1\over {2 \kappa^2}}
  {R} -  {1\over 2} \partial_{M} \Phi
  {\partial^{M} {\bar \Phi}} - {{\lambda}\over {4}} (\Phi {\bar \Phi} 
- \eta^2)^2- {\Lambda}\right)},
\label{5D-complex-scalar-action}
\eeq
with $M,N,=0,...4$ and $\kappa^2 = M_{P}^{-3}$, where  $M_{P}$ 
is the $5d$ reduced Planck mass, and $\Lambda$ denotes the
five-dimensional cosmological constant.  We now review the results 
of \cite{BP-SP-V-1}.

\subsection{The flux vacua}

To begin, we shall constrain the magnitude of the scalar field
to lie at $|\Phi|=\eta$, leading to the effective action
\beq
S=\int{d^5 x \sqrt{- g} \left({1\over {2\kappa^2}} R - {1\over 2} \eta^2
  \partial_{M} \theta \partial^{M} \theta - {\Lambda}\right)},
\label{5D-action}
\eeq
where $\theta$ is the phase of $\Phi$.
The equations of motion for this model are
\beq
\partial_M \left(\sqrt{-g} \partial^M \theta \right)=0~,
\label{scalar-eom}
\eeq
\beq
R_{AB} - {1\over 2} g_{AB} R
= \kappa^2 T_{AB}~,
\label{Einstein-eq-5D}
\eeq
where
\beq
T_{AB} = \eta^2 \left(\partial_A \theta \partial_B \theta - {1\over 2}
g_{AB} \partial_{M} \theta \partial^{M} \theta \right)
- g_{AB} \Lambda~
\eeq
is the energy momentum tensor.  We will look for a solution of the form
\beq
ds^2= g_{MN} dx^M dx^N = g_{\mu \nu} d x^{\mu}
d x^{\nu} +  g_{yy}(x^{\mu}) dy^2,
\label{5D-metric}
\eeq
where $\mu, \nu = 0,1,2,3$ denote the $4d$ coordinates, and 
the compact extra dimension is parameterized by the coordinate 
$0 \leq y < 2 \pi$.  We are interested in the case
\beq
g_{yy}(x^{\mu}) = L^2 = const.~,
\eeq
i.e., in solutions with the extra dimension stabilized
at a constant circumference $2\pi L$.  We shall also require 
that the four-dimensional slices are described by a spacetime of 
maximal symmetry whose scalar curvature is given by $R^{(4)} = 12
H^2$, with $H^2$ negative for the anti-de
Sitter (AdS) case.  

The solutions to the equations of motion for the scalar field 
Eq.  (\ref{scalar-eom}) which are compatible with maximal spacetime 
symmetry are given by
\beq
\theta (x^M) = n~y~.
\label{thetax5}
\eeq
The change of axion phase $\theta$ around the compact dimension must be an
integer multiple of $2\pi$, and hence the various flux vacua are parameterized by
the integer $n$.

With these assumptions we arrive at the Einstein equations
\beq
3 H^2 = \kappa^2 \left( {{n^2 \eta^2}\over{2 L^2}} +
\Lambda \right),
\eeq
\beq
6 H^2  = - \kappa^2 \left({{n^2\eta^2}\over {2 L^2}} - \Lambda\right),
\eeq
which fix the values of $H$ and $L$ in terms of $n$ and the parameters
of the original Lagrangian according to
\beq
L^2 = - {{3 n^2 \eta^2 }\over {2 \Lambda}},
\label{L2}
\eeq
\beq
H^2 = {{2 \kappa^2 \Lambda} \over {9}}~.
\label{H2}
\eeq 

We thus conclude from Eq.  (\ref{L2}) that flux vacua exist for 
$S^1$ compactifications provided the $5d$ cosmological constant 
is negative ($\Lambda < 0$).  Eq.~(\ref{H2}) then indicates 
that these compactifications always yield a $4d$ anti-de Sitter
spacetime, enumerated by the integer $n \neq 0$.
Furthermore, it can be easily shown by studying the $4d$ effective action
associated with these models \cite{BP-SP-V-1}, that the above
solutions are perturbatively stable.  
We will now investigate the possibility that there exist
non-pertubative instabilities of this geometry\footnote{Flux-changing instantons \cite{BP-SP-V-1} have previously been shown to exist.  Their relation to the bubble of nothing presented here will be discussed below.}, similar to the pure
Kaluza -- Klein bubble of nothing presented in the previous section.

\subsection{The bubble of nothing as a de Sitter brane}

A particularly useful description of our $AdS_4 \times S^1$ compactification 
is given by the following metric which covers a Rindler-like portion 
of the $4d$ spacetime shown in Fig.~(\ref{confdiag23}) below:
\beq
ds^2 =  dr^2 +  H^{-2} \sinh^2 (H r) (-dt^2 + \cosh^2t\,\,d \Omega_2^2) + L^2 dy^2~.
\label{RindlerAdS}
\eeq

\begin{figure}[h] 
\centering
\includegraphics[width=3.0cm]{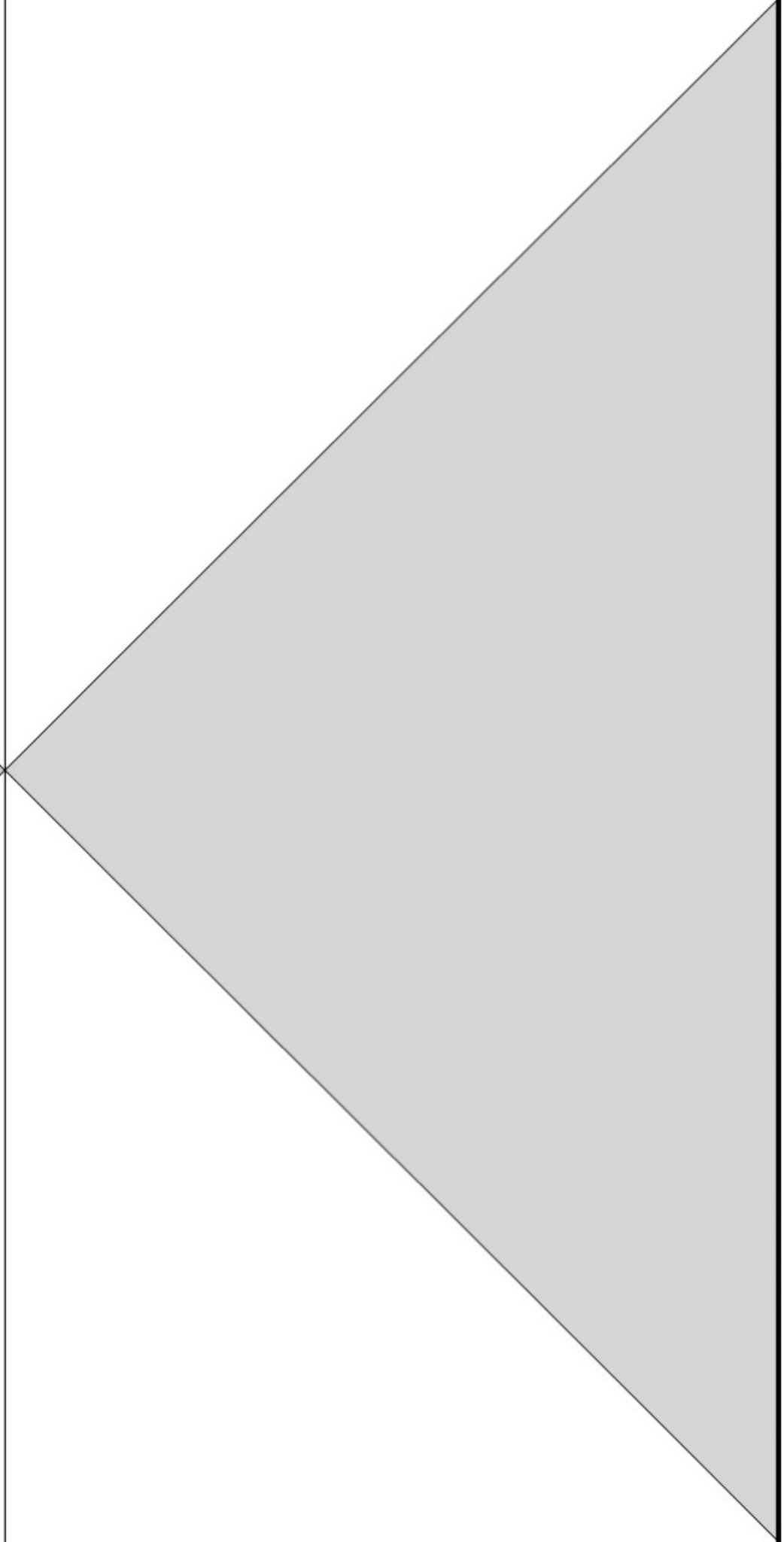}\hspace{4cm}
\includegraphics[width=3.0cm]{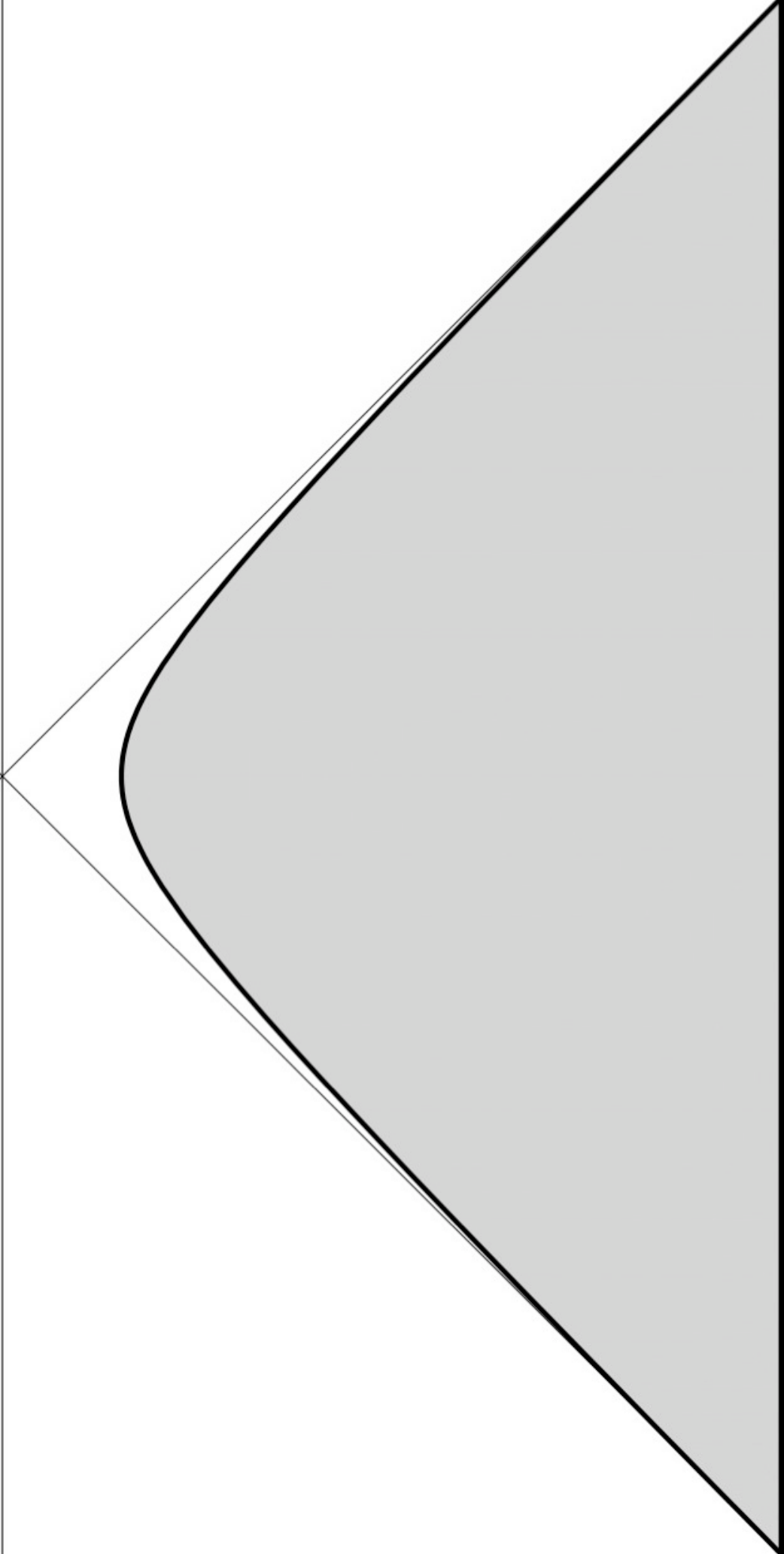}
\caption{Left:  A conformal diagram for $AdS_4$.  
An $S^2$ is suppresed at each point except the left vertical line.  The 
coordinates used in Eq.~(\ref{RindlerAdS}) cover the shaded region.  
Right:  The spacetime only exists outside (shaded region) the bubble 
wall, which respects the de Sitter slicing of
$AdS_4$.} 
\label{confdiag23}
\end{figure}

Comparing with the right hand illustration in Fig.~(\ref{confdiag23}), 
consider a metric of the form
\beq
ds^2 =  dr^2 + B^2 (r) (-dt^2 + \cosh^2t\,\,d\Omega_2^2) + r^2C(r)^2 dy^2
\label{ansatz}
\eeq
with boundary conditions
\beq
B(r) \rightarrow \ell ~~~~~~~~~ C(r)\rightarrow  1
\eeq
for $r \rightarrow 0$, and

\beq
{{\partial_r B(r)}\over{B(r)}} \rightarrow  
H\coth(H r) ~~~~~~~~ r C(r) \rightarrow  L
\eeq
in the limit $r \rightarrow \infty$.  

It is clear from this description that such a solution, if it exists, 
would have the appropriate asymptotics as $r \rightarrow \infty$ 
to match to the flux compactification solution Eq.~(\ref{RindlerAdS}) at the conformal 
boundary, where we must additionally impose that the 
axion $\theta$ approach the form Eq.~(\ref{thetax5}).
Looking at the boundary conditions imposed above, one can see that
the metric in Eq.~(\ref{ansatz}) has the same structure at 
$r \rightarrow 0$ as Witten's solution  discussed
previously in Eq.~(\ref{wittenmetric});  we have a compelling ansatz for a bubble of 
nothing in this flux compactification.

Because the compact dimension in this solution closes smoothly 
at $r = 0$, one must introduce some dynamical object that is able to 
resolve the flux singularity on the surface of the bubble.  We can
accomplish this simply by examining our original Lagrangian for the complex scalar
field in Eq.~(\ref{5D-complex-scalar-action}).  The 
problem arises only if one insists on keeping the modulus of the scalar
field finite near the surface of the bubble.  The divergence of gradient energy is cured
by allowing the scalar modulus to relax along the radial
direction in such a way that it vanishes at the tip of the cigar-like
geometry.  This is the same regularization found on a 
global string (i.e.  codimension two) solitonic solution associated with 
a complex scalar field.  

Singular bubbles of nothing in $AdS_5\times S^5/\mathbbm{Z}_k$ were
constructed in \cite{Horowitz:2007pr}, yielding a bubble
surface charged with respect to the stabilizing flux.  The 
bubble, located where an $S^1$ fiber degenerates, 
is singular due to the required de Sitter D3 branes
smeared on the $\mathbbm{C}P^2$ base.  Our aim is to find non-singular
solutions using a solitonic rather than smeared point-like sources.

As pointed out in \cite{BP-SP-V-1}, our Lagrangian admits such
solitonic solutions describing 2-branes {\it charged} with respect to the axion 
$\theta$.  We therefore conjecture that one should identify the 
solution described above, the bubble of nothing within our
$AdS_4\times S^1$ flux compactification, as a de Sitter solitonic 2-brane in 
a $5d$ AdS spacetime.  We prove in the next section that one
can indeed find such smooth bubble geometries in our model
by numerically solving the equations of motion using the 
ansatz described above.

Solutions for de Sitter branes as topological defects have been previously 
discussed in \cite{Cho-V}, whose numerical solutions 
introduced many characteristics found in our
bubble of nothing.  Here we give a different interpretation for 
these spacetimes in the context of flux compactifications.  
We illustrate such a bubble of nothing in Fig.~(\ref{bubble2}) below.

\begin{figure}[h] 
\centering
\includegraphics[width=8.0cm]{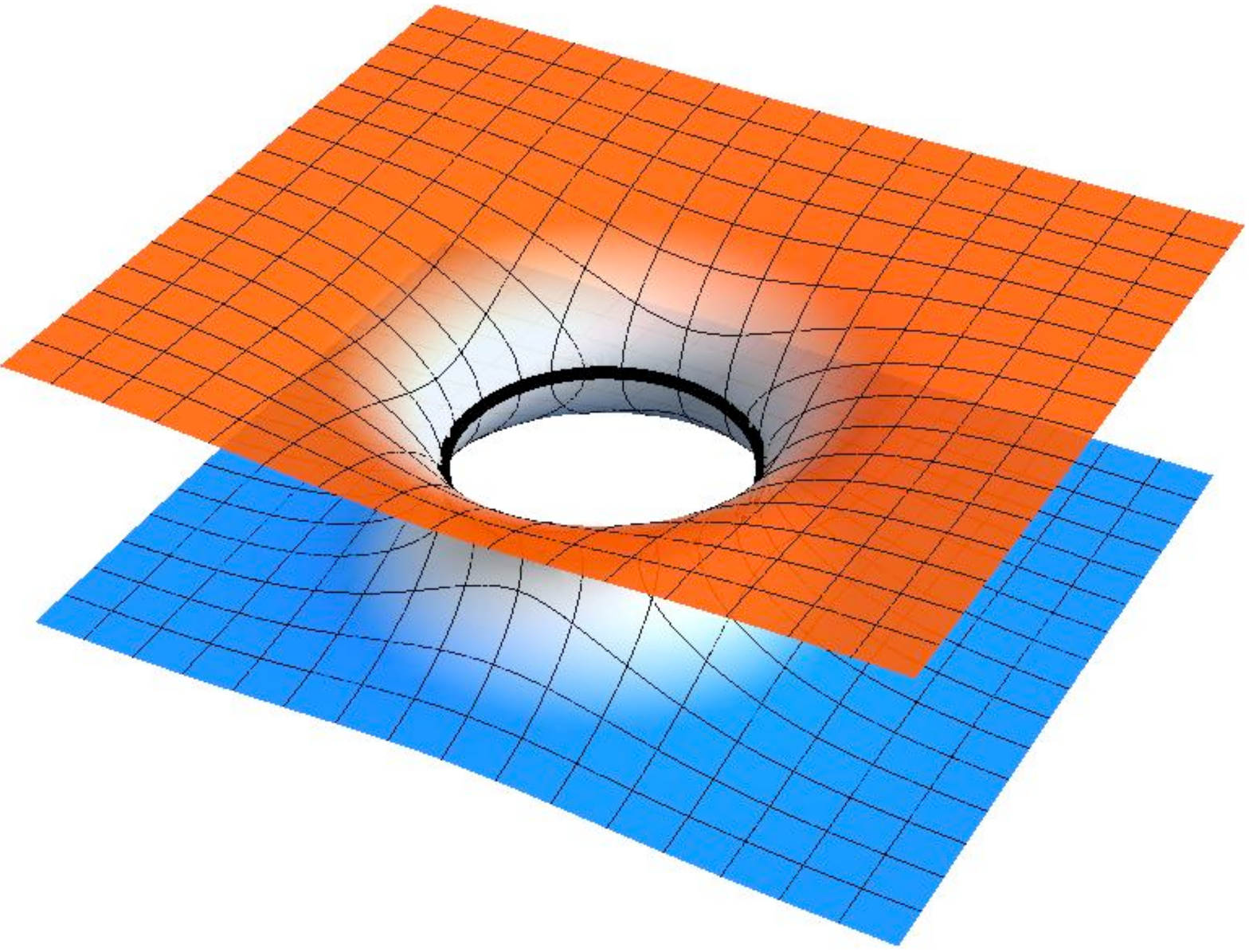}\hspace{1cm}
\includegraphics[width=7.0cm]{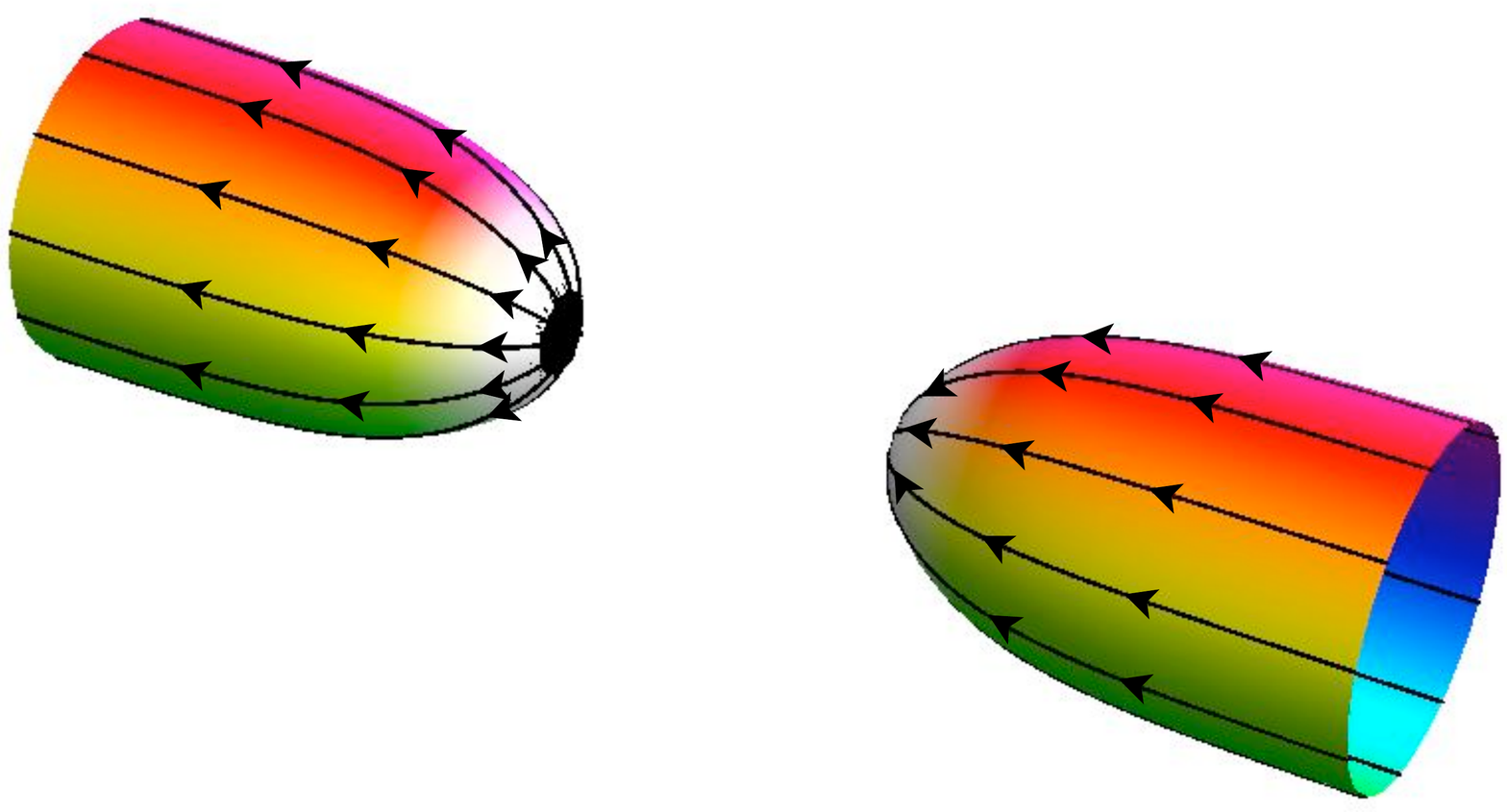}
\caption{An axionic bubble of nothing.  As in Witten's solution, the bubble wall lies where the $S^1$ extra dimension 
degenerates.  In this case, the wall is charged under an axionic
phase.  Hue and saturation represent the scalar phase $\theta$, and modulus $f$, respectively.  
The core of the solitonic 2-brane has an ill-defined phase, and is shown in black.  The left-hand picture shows two large dimensions, with the 
apparent vertical separation representative of the KK radion.  The 
right-hand picture shows the full KK extra dimension, but only one 
of the four large dimensions.} 
\label{bubble2}
\end{figure}

\section{The Bubble Solution}

We begin by considering the action
\beq
S=\int{d^5 x \sqrt{- g} \left( {1\over {2\kappa^2}}
  R -  {1\over 2} \partial_{M} \Phi
  {\partial^{M} {\bar \Phi}} - {{\lambda}\over {4}} (\Phi {\bar \Phi} 
- \eta^2)^2- \Lambda\right)},
\eeq
with metric
\beq
ds^2 =  dr^2 + B^2 (r) (-dt^2 + \cosh^2(t) d\Omega_2^2) + r^2C(r)^2 dy^2~,
\eeq
and scalar field
\beq
\Phi(x^M) = f(r)\mbox{e}^{i\theta(y)} = f(r) \mbox{e}^{i n y}~.
\eeq
The $O(3,1)\times U(1)$ symmetry of the bubble of nothing is easily 
seen within this ansatz.  We arrive at the equations
\beq
R_{MN} - {1\over 2} g_{MN} R = \kappa^2 T_{MN}~,
\eeq
and
\beq
\partial_M(\sqrt{-g} \partial^M \Phi) - \sqrt{-g} \lambda
\Phi (|\Phi|^2 - \eta^2) = 0~,
\eeq
having denoted
\beq
T_{MN} = \partial_M \Phi \partial_N \bar \Phi + g_{MN} \left(- {1\over 2}
  \partial_P \Phi \partial^P \bar \Phi - {{\lambda} \over {4}} (\Phi
 \bar \Phi - \eta^2)^2 - \Lambda\right)~.
\eeq
The Einstein tensor then becomes
\begin{eqnarray}
G^{t}_{t} &=&  - {1\over {B^2}} + {{2 B'}\over {r B}} +
{{B'^2}\over {B^2}} + {{2C'}\over {r C}} +  {{2B' C'}\over {B C}}
  + {{2B''}\over {B}} + {{C''}\over {C}}  \nonumber  \\ \nonumber \\
G^{r}_{r} &=& -{{3}\over {B^2}} + {{3B'}\over {r B}} + {{3
    B'^2}\over {B^2}} + 
{{3 B' C'}\over {B C}}  \nonumber  \\ \nonumber \\
G^{y}_{y} &=& -{{3}\over {B^2}} + {{3B'^2}\over {B^2}} + {{3B''}\over {B}}
\end{eqnarray}
and the energy momentum tensor is given by
\begin{eqnarray}
T^{t}_{t} &=& - {1\over 2} f'^2 - {{n^2f^2}\over {2 r^2 C^2}} 
- {{\lambda} \over {4}} (f^2 - \eta^2)^2 - \Lambda \nonumber  \\ \nonumber \\
T^{r}_{r} &=&  {1\over 2} f'^2 - {{n^2f^2}\over {2r^2 C^2}} 
- {{\lambda} \over {4}} (f^2 - \eta^2)^2 - \Lambda \nonumber \\ \nonumber \\
T^{y}_{y} &=& - {1\over 2} f'^2  + {{n^2f^2}\over {2r^2 C^2}} 
- {{\lambda} \over {4}} (f^2 - \eta^2)^2 - \Lambda \nonumber~.
\end{eqnarray}
The equations of motion are then any three of the four equations
\begin{eqnarray}
- {1\over {B^2}} + {{2 B'}\over {r B}} +
{{B'^2}\over {B^2}} + {{2C'}\over {r C}} +  {{2B' C'}\over {B C}}
  + {{2B''}\over {B}} + {{C''}\over {C}} &=& \kappa^2 
\left(- {1\over 2} f'^2 - {{n^2f^2}\over {2 r^2 C^2}} 
- {{\lambda} \over {4}} (f^2 - \eta^2)^2 - \Lambda \right) \nonumber  \\ \nonumber \\
-{{3}\over {B^2}} + {{3B'}\over {r B}} + {{3
    B'^2}\over {B^2}} + 
{{3 B' C'}\over {B C}} &=&  \kappa^2 
\left({1\over 2} f'^2 - {{n^2f^2}\over {2 r^2 C^2}} 
- {{\lambda} \over {4}} (f^2 - \eta^2)^2 - \Lambda \right)\nonumber \\ \nonumber \\
-{{3}\over {B^2}} + {{3B'^2}\over {B^2}} + {{3B''}\over {B}} &=& \kappa^2 
\left( - {1\over 2} f'^2  + {{n^2f^2}\over {2 r^2 C^2}} 
- {{\lambda} \over {4}} (f^2 - \eta^2)^2 - \Lambda \right) \nonumber\\
f'' +\left( 3 {{B'}\over {B}} + {{C'}\over {C}} + {1\over r}
\right) f' &=& {{n^2f} \over {C^2 r^2}} + \lambda f (f^2 - \eta^2)~.\nonumber
\end{eqnarray}
Because our ansatz is explicitly time-independent, the Lorentzian and Euclidean
solutions are trivially related.  

\subsection{Asymptotic Solution of the full equations}
Before numerically solving the near-bubble region, we determine
the asymptotic values of all functions, which can be done exactly.
The solution below exhibits the expected backreaction on the scalar
modulus, the KK radion, and the vacuum energy density.
\beq
\label{phiasymsol}
\Phi(x) = f_{\infty} e^{i n y}~,
\eeq
\beq
\label{metricasymsol}
ds^2 =  dr^2 + {1\over {H^2}} \sinh^2(H r) 
(-dt^2 + \cosh^2(t) d\Omega_2^2) + L^2 dy^2~,
\eeq
where,
\beq
f_{\infty}^2 = \eta^2 - {{n^2}\over {\lambda L^2}} = {{2 \eta^2}\over
  {5}} \left(1 + {3 \over 2}\Delta\right)~,
\eeq
\beq
L^2 = -{{3n^2 \eta^2}\over {4 \Lambda}}\left(1 + \Delta \right)~,
\eeq
\beq
H^2 = -{{4 \kappa^2 \Lambda}\over {15}} \left({{{2\over 3} + \Delta}\over {1+ \Delta }}\right)~,
\eeq
and we have introduced the parameter $\Delta$,
\beq
 \Delta = \sqrt{1 + {{20
      \Lambda}\over {9 \eta^4 \lambda}}}~.
\eeq
Note that $\Delta \to 1$ in the limit 
$\lambda \rightarrow \infty$, so we recover the pure flux
compactification geometry described in previous section (See
Eqs.  (\ref{L2}) - (\ref{H2})).

\subsection{Near core expansion}

We can Taylor expand the equations of motion about
the tip of the cigar.  This leaves two unknown boundary conditions, 
which we will use as shooting parameters.  For $n = 1$ they are $\ell$ and $f^\prime_0$.  
The remaining terms are then completely specified:

\begin{eqnarray}
B(r) &=& \ell + \left(\frac{1}{2\ell} - \frac{\kappa^2\ell 
(\eta^4 \lambda + 4 \Lambda)}{24}\right) r^2 + ...\\
C(r) &=& 1 +  \left(-\frac{1}{2\ell^2} + \frac{\kappa^2 
\left(\eta^4 \lambda + 4 \Lambda - 12 {f^\prime_0}^2\right)}{72}\right) r^2 + ...\\
f(r) &=& f^\prime_0 r\, +\,\, ...
\end{eqnarray}

\begin{figure}[t] 
\centering
\includegraphics[width=8.0cm]{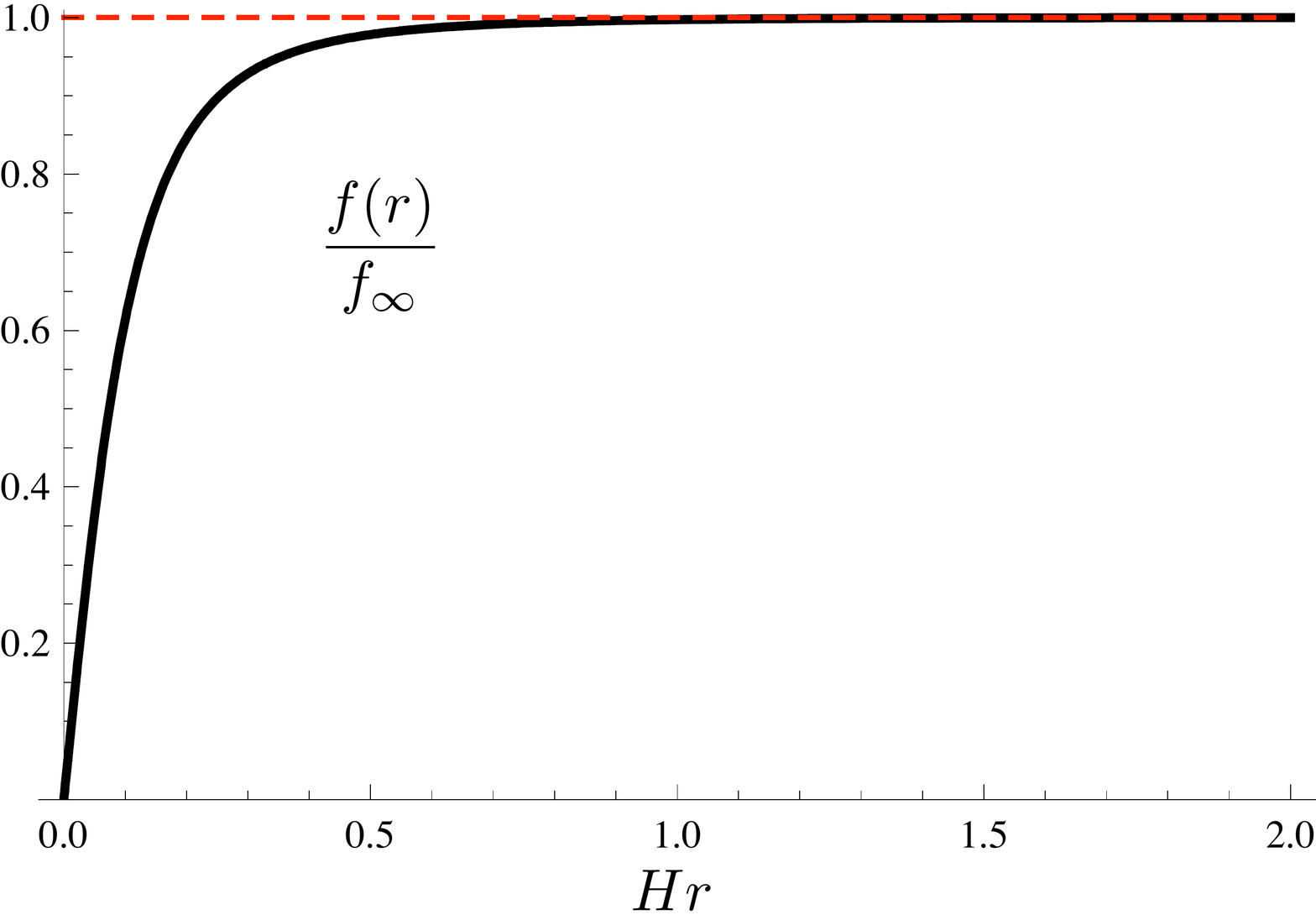}
\includegraphics[width=8.0cm]{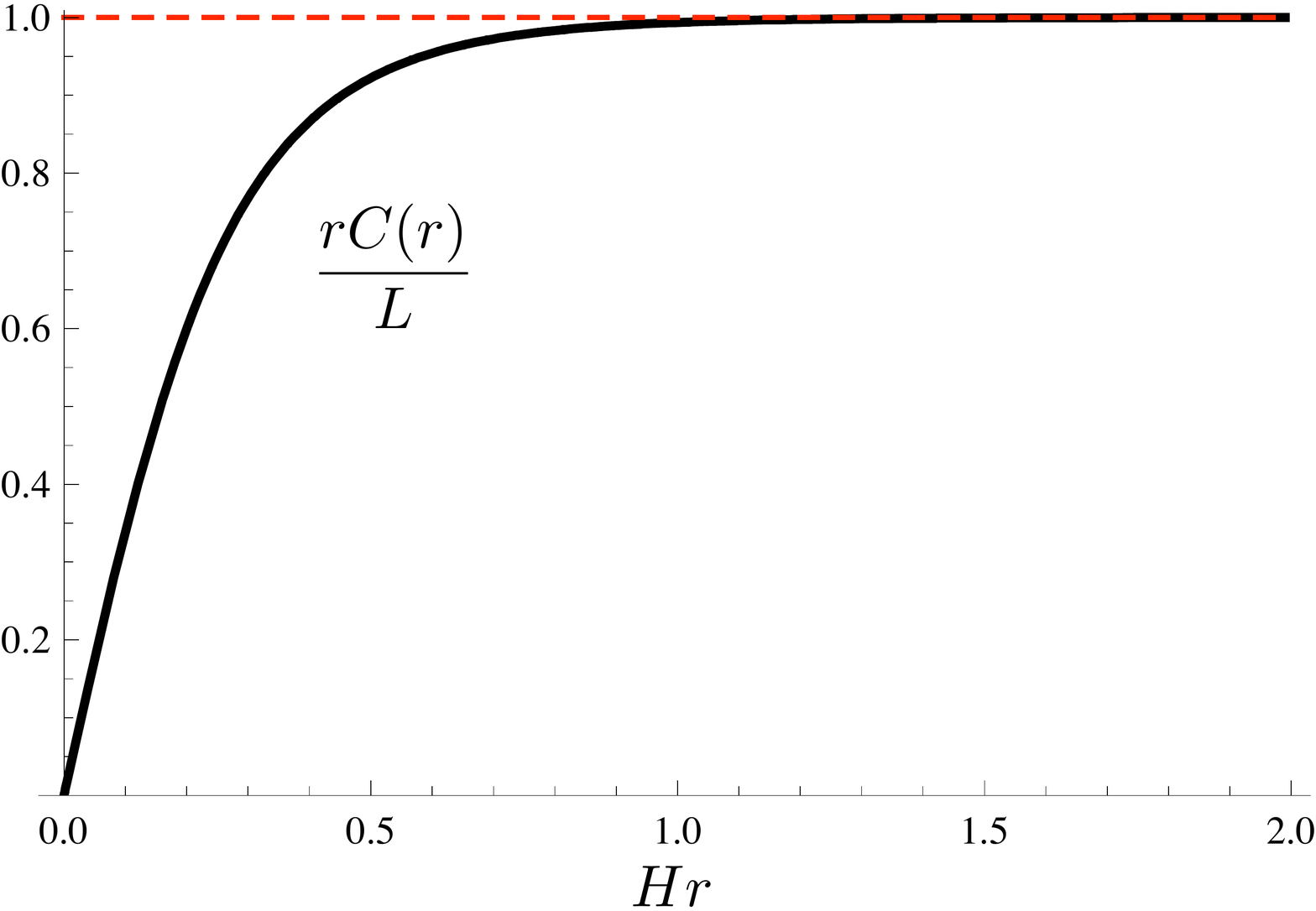}
\includegraphics[width=8.0cm]{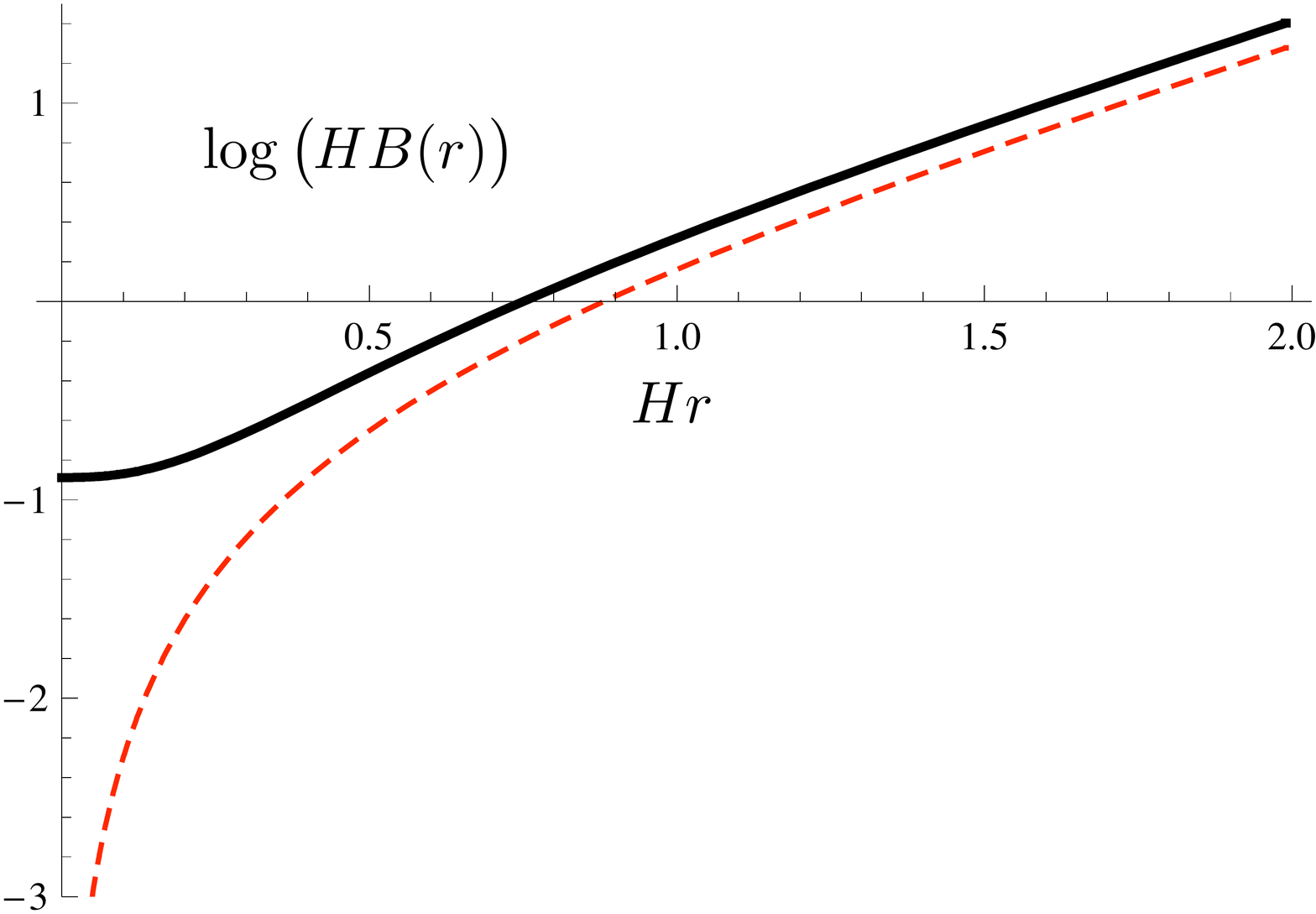}
\includegraphics[width=8.0cm]{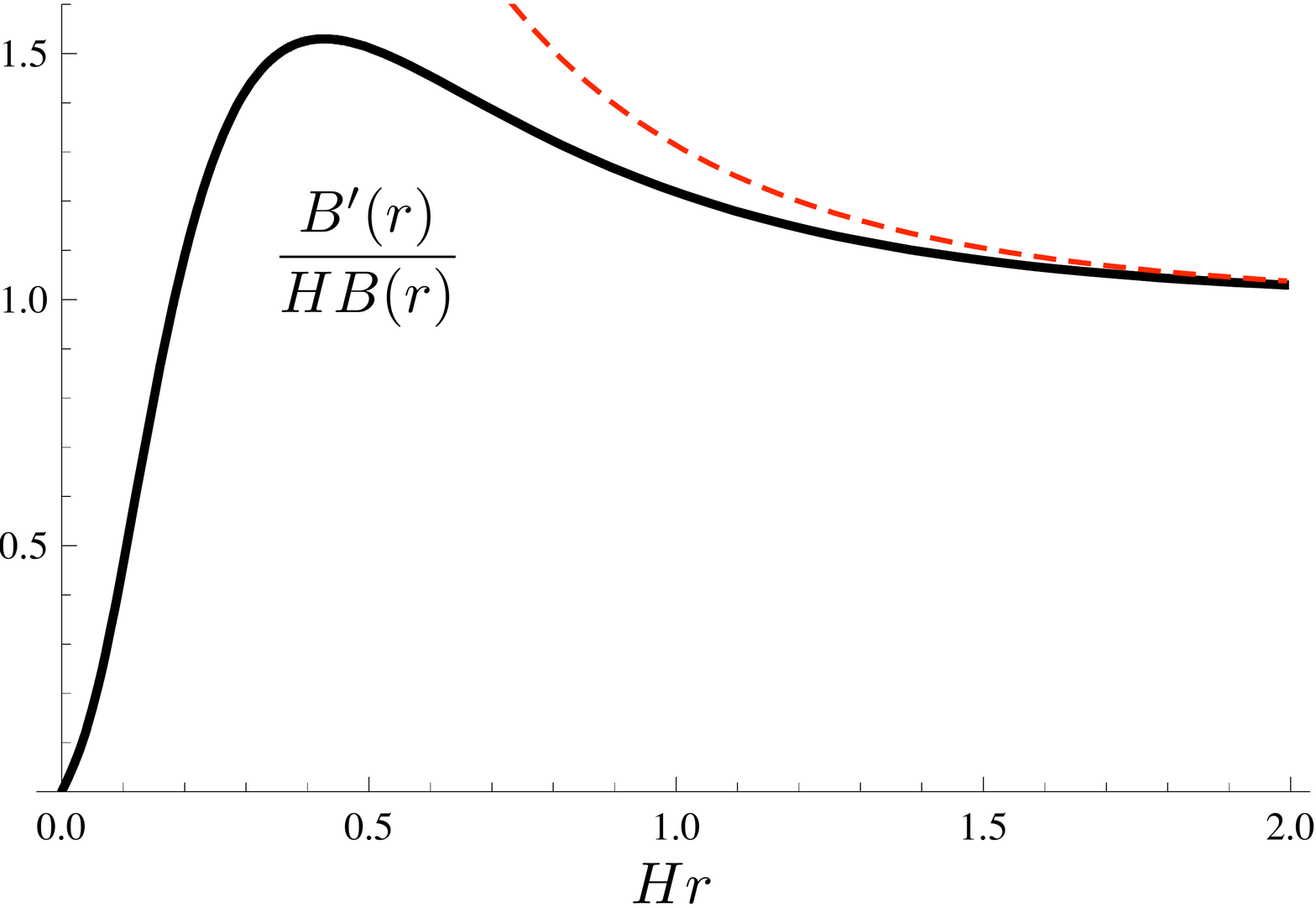}
\caption{The numerical solutions for the scalar modulus (top left), 
the KK radion modulus (top right), 
the de Sitter slice radius (bottom left), and
its derivative (bottom right).  We express 
$r$ in units of the AdS radius $1/H$.  Dashed red lines 
represent the pure compactification solution of Eqs.
(\ref{phiasymsol} - \ref{metricasymsol}), which shares the conformal 
boundary with the bubble solution.} 
\label{numericalsoln}
\end{figure}

We then numerically integrate the equations of motion outward from $r= 0$ to obtain
the functions pictured in Fig.~(\ref{numericalsoln}).  The numerical values for the parameters 
used are
\beq
n = 1\qquad
\Lambda = -(0.347 M_P)^5\qquad
\eta = (0.630 M_P)^{3/2}\qquad
\lambda = (0.995 M_P)^{-1}~,
\eeq
which give the solution
\beq
H = 0.0332 M_P\qquad
L = (0.118 M_P)^{-1}\qquad
\ell = (0.0806 M_P)^{-1}\qquad
f^\prime_0 = (0.456 M_P)^{5/2}~,
\eeq
where $M_P = \kappa^{-2/3}$ is the $5d$ reduced Planck mass.  The Euclidean action of this solution is given by
\beq
S_B = S_E\left[\mbox{bubble}\right] - S_E\left[\mbox{compactification}\right] \approx 7.4 \times 10^4~,
\eeq
where the numerical solution was matched to the background in a box of circumference $25.566 H^{-1}$.
Our chosen parameter values are rather generic, but we expect a large
range of solutions to exist.  In particular, we can smoothly deform our parameters and corresponding solution to match the bubble of Witten.

\begin{figure}[t] 
\centering
\includegraphics[width=8.0cm]{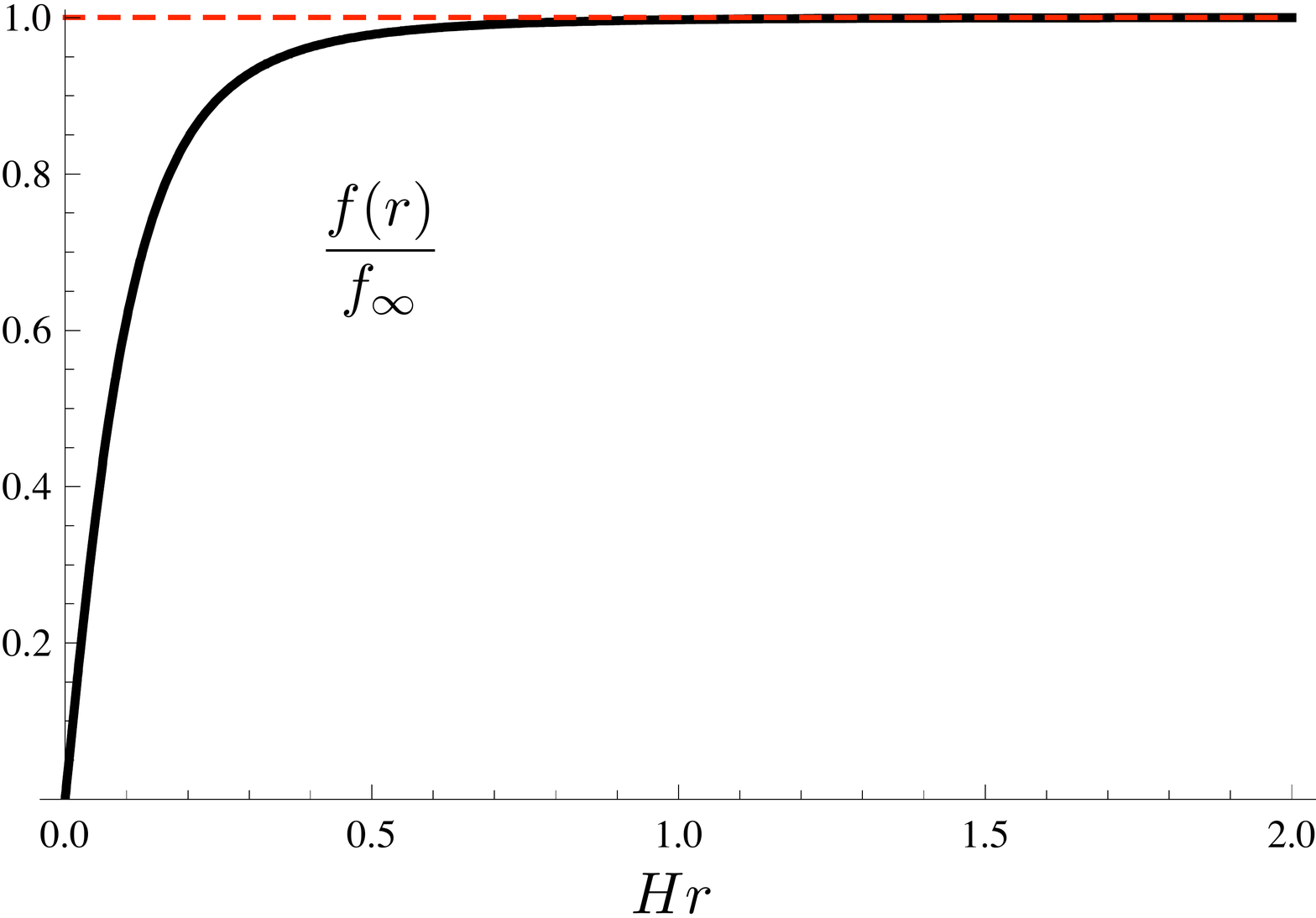}
\includegraphics[width=8.0cm]{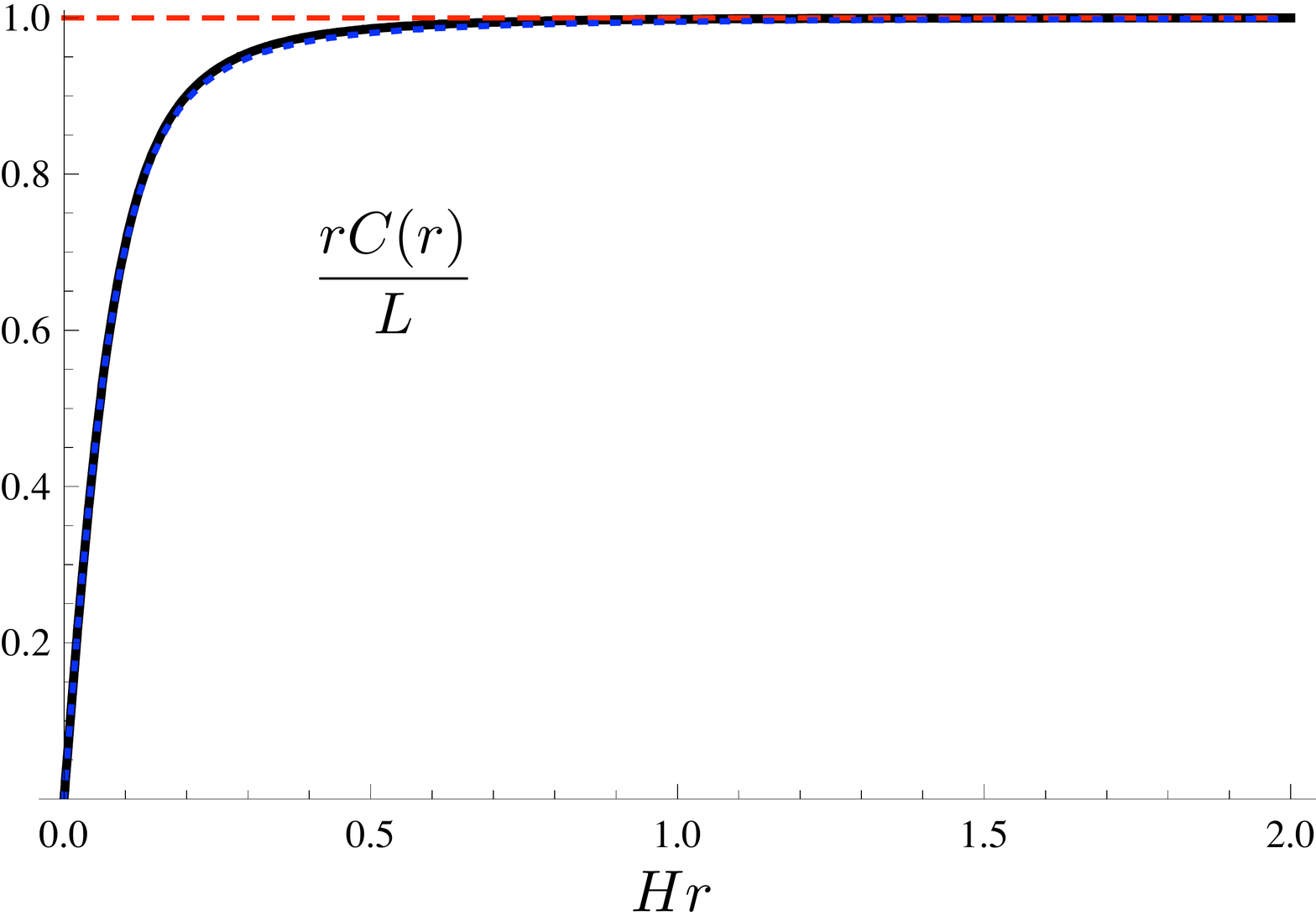}
\includegraphics[width=8.0cm]{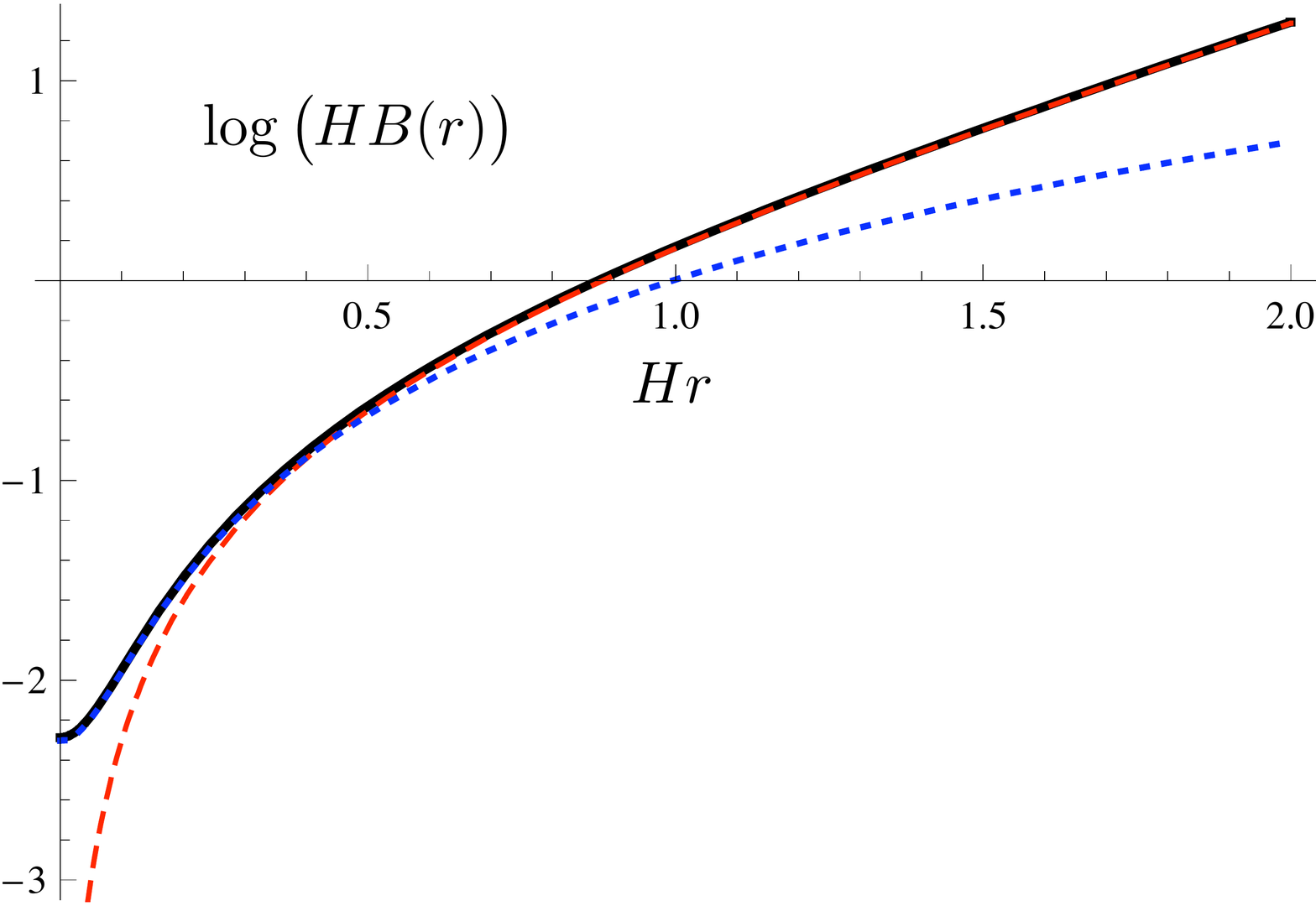}
\includegraphics[width=8.0cm]{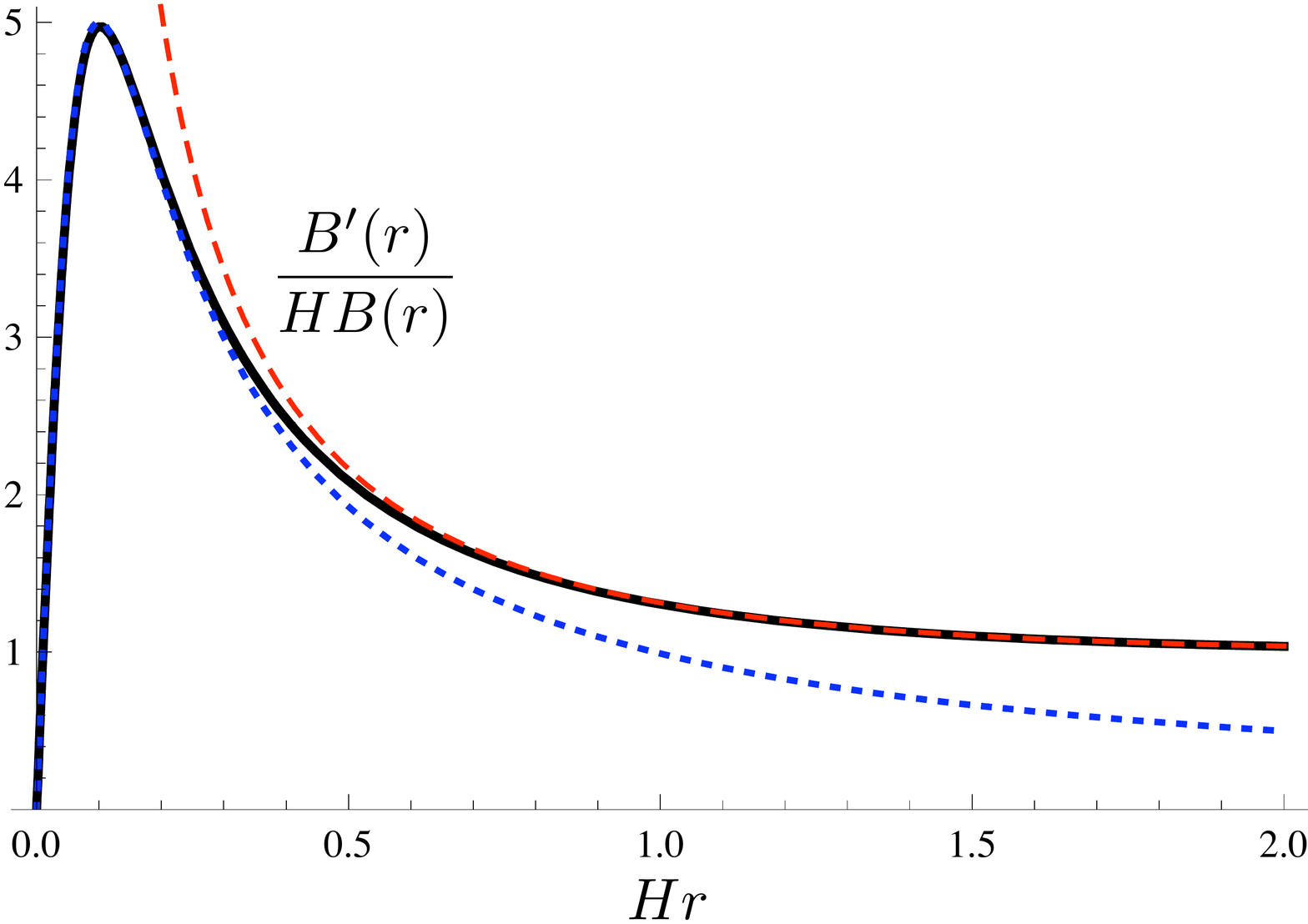}
\caption{Parameters were chosen to demonstrate the smooth embedding of Witten's bubble in a mildly AdS flux compactification.  Dotted blue lines represent Witten's solution of identical KK circumference $2\pi L$.  Dashed red lines 
represent the pure flux compactification solution, as before.  Near the bubble, the solution resembles that of Witten, while on scales larger than $H^{-1}$, the solution approaches the AdS compactification.} 
\label{wittennumericalsoln}
\end{figure}

This is shown in the solution Fig.~(\ref{wittennumericalsoln}), where the chosen parameters are
\beq
n = 1\qquad
\Lambda = -(0.224 M_P)^5\qquad
\eta = (0.422 M_P)^{3/2}\qquad
\lambda = (4.50 M_P)^{-1}~,
\eeq
which give the solution
\beq
H =  0.010 M_P\qquad
L = (0.100 M_P)^{-1}\qquad
\ell = (0.0985 M_P)^{-1}\qquad
f^\prime_0 = (0.179 M_P)^{5/2}~.
\eeq
The Euclidean action of this solution is given by
\beq
S_B = S_E\left[\mbox{bubble}\right] - S_E\left[\mbox{compactification}\right] \approx 6.4 \times 10^4~,
\eeq
which is similar to the analogous action for Witten: $S_B = 2\pi^3L^3/\kappa^2 \approx 6.2\times 10^4$.  As seen in Fig.~(\ref{wittennumericalsoln}), the solution resembles that of Witten on scales smaller than $H^{-1}$, but then asymptotes to the AdS compactification far from the bubble.

\section{Conclusions}

We have shown that a simple $AdS_4\times S^1$ flux compactification exhibits an 
instability to the nucleation of bubbles of nothing.  The key
feature of this geometry is the flux through the 
compactification cycle at the conformal boundary, which demands a source wherever 
the cycle degenerates.  Appropriately, the field theoretic model considered
has solitonic 2-brane solutions which are charged with respect to the axionic flux stabilizing the extra dimension.  One can construct a bubble of nothing in this context as the 
spacetime sourced by an ``inflating" solitonic 2-brane whose
worldvolume is given by a codimension-two de Sitter space
(the surface of the bubble).  We have obtained numerical examples
of these bubbles which asymptotically match the
pure compactification geometry.  

We have presented our construction using the simplest field 
theory that possesses the minimal ingredients for the scenario 
to work, but we believe this kind of instability should be 
a generic feature of other non-supersymmetric flux compactifications.  
In particular, it will be straightforward to implement these 
ideas in a model based on a $6d$ Einstein -- Maxwell theory, where one 
finds a sizable landscape \cite{ME-6d, BP-SP-V-1} of $dS_4$, $M_4$ and
$AdS_4$ vacua.  As explained in \cite{BP-SP-V-1}, this theory
possesses 2-brane solutions that could be used to construct 
instantons which interpolate between vacua of differing flux numbers.
One can then extend this model to an Einstein -- Yang -- Mills -- Higgs theory
\cite{Cremmer}.  The new degrees of freedom incorporated in this model 
allow for smooth codimension-three solitonic 2-branes.  One should then be
able to find a magnetically charged de Sitter 2-brane
with the correct asymptotics, similar to those in 
\cite{Cho-V-2}.  This type of bubble represents a significant departure from existing constructions, since the degenerating sphere is of dimension $> 1$.  One consequence of this topology is that spin structure will not play a role in excluding the instability.  We are currently studying such solutions and will present our results in a separate publication.

An alternate perspective of the solutions presented here is
as the critical case of a flux-changing instanton whose 
final vacuum has zero flux.  Typical flux-changing
transitions do not exhibit large backreaction on the
spacetime geometry, provided one considers the nucleation of a 
single-charge brane, i.e., small relative changes in flux.  On the other
hand, the bubble of nothing corresponds to the extreme case, 
where the brane acts as a sink for {\em all} the flux found in the
asymptotic compactification.  It is not surprising that the
effect on the geometry is drastic, since a zero-flux vacuum is
unstable to collapse.  One can see this in our model
by examining the effective $4d$ theory for the KK radion,
which in the absence of flux is parameterized solely by the contribution
from the negative bulk cosmological constant.  The resulting tachyonic potential is
divergent as one approaches zero-size compact dimension 
(See \cite{BP-SP-V-1}).  This suggests that one should be
able to understand the bubble of nothing from a
purely $4d$ point of view by including the field $f$ as a
new scalar degree of freedom in the $4d$ effective theory.  Such a
solution may appear singular from the four-dimensional perspective, 
but this is just an artifact of the dimensional reduction, 
much like the case of Witten's bubble \cite{Garriga}.

The bubbles of nothing considered here affect many
otherwise stable flux vacua.  One should consider 
this instability as a new decay channel whose end result is
a type of terminal vacuum, a vacuum without a classical spacetime.
This is clearly relevant for any program undertaking the
assigning of probabilities to different vacua in the
landscape, the so-called measure problem.  (See for example,
\cite{measure} and references therein.)

\section{Acknowledgments}
We would like to thank Jaume Garriga, Oriol Pujolas and Alex Vilenkin
for very helpful conversations and discussions.  J.J.B.-P.  would like
to thank the Theory Division at CERN for its hospitality and support
in the early stages of this work.  J.J.B.-P.  is supported 
in part by the National Science Foundation under grant 06533561.  B.S.  
is supported in part by Foundational Questions Institute grant RFP2-08-26A.

\end{document}